\documentclass[showpacs,twocolumn,prl]{revtex4}
\usepackage{amsmath,graphicx}

\begin{document}

\title{Dissipation  and supercurrent  fluctuations  in a diffusive NS ring}
\author{B. Dassonneville$^{1}$, M. Ferrier$^{1}$,  S. Gu\'eron$^{1}$ and H.Bouchiat $^{1}$ }

\affiliation{$^{1}$ LPS, Univ. Paris-Sud, CNRS, UMR 8502, F-91405 Orsay Cedex, France}

\begin{abstract}
 A mesoscopic hybrid Normal/Superconducting (NS) ring is characterized by a dense  Andreev spectrum  with a flux dependent minigap.  To probe the dynamics  of such a ring  we measure its linear response to a high frequency flux, in a wide frequency range, with a multimode superconducting resonator. We find that the current response contains, beside the well known dissipationless  Josephson contribution, a large dissipative component. At high frequency  compared to the minigap and low temperature we find that the dissipation is due to transitions across the minigap. In contrast, at lower frequency there is a range of temperature for which dissipation is caused predominantly by the relaxation of the Andreev states' population. %{\it and for which the response is proportional to the  square of the average square of the single level current?? dur dur... *}..
  This dissipative response, related via the fluctuation dissipation theorem to a non intuitive
 zero  frequency  thermal noise of  supercurrent, is characterized by a phase dependence dominated by its second harmonic, as predicted long ago  \cite{limpitsky,martin96} but never observed so far.%in the context of SNS junctions  \cite{limpitski, avouris, yeyati} as well as persistent currents in normal rings \cite {buttiker85, trivedi88}.
\end{abstract}

\maketitle

A  phase coherent non superconducting  conductor (N) connected to two superconductors (an SNS junction) gives rise to the formation of Andreev states (AS) which are coherent superpositions of electron and hole states  confined in the N metal and carry the Josephson-supercurrent \cite{kulik}. They strongly  depend on the  phase difference $ \varphi $ between the two superconductors. The quasi-continuous Andreev spectrum of a diffusive metallic wire exhibits a phase modulated induced gap, the minigap, which  closes at odd multiples of $\pi$ (for perfectly transmitting contacts), and can be approximated by $2E_g(\varphi)= 2E_g(0)|cos\varphi/2|$ \cite{heikkila,spivak,blatter}. In the case of a wire longer than the superconducting coherence length, $E_g(0) \simeq 3E_{Th}$, with $E_{Th}$ the Thouless energy $\hbar/\tau_D$, and $\tau_D$ the diffusion time across the N wire.
 Whereas most  investigations of  diffusive SNS junctions rely on non linear transport measurements of current voltage curves,  switching current and ac Josephson effect \cite{dubos,lehnert}, only few experiments probe the Andreev spectrum at equilibrium in a phase biased  configuration with NS rings threaded by an Aharonov Bohm  flux. These include the tunnel spectroscopy of the minigap \cite{lesueur} and the measurement of the flux dependent Josephson supercurrent using SQUID\cite {illichev} or Hall probe magnetometry \cite{strunk}.  Beyond probing the equilibrium AS spectrum in a {\it static} magnetic  flux, the investigation of the {\it dynamics} associated to  this spectrum, a  far more complex question, has been addressed experimentally only recently\cite{chiodi2011}. Theoretically, it was predicted that, in contrast to tunnel Josephson junctions \cite{noiseJJ} and because of the smallness of the induced gap,  SNS junctions should exhibit low frequency supercurrent  fluctuations at equilibrium \cite{martin96}. According to the fluctuation dissipation theorem,  in the linear response regime, such equilibrium fluctuations lead to a  dissipative current under an ac flux excitation \cite{kulikintau,virtanen}. 
In this Letter, we present  the  linear current response  of a phase biased NS ring and account for both the non dissipative  component and  the more surprising dissipative one, over a wide frequency range. We identify two fundamental processes leading to dissipation, the microwave-induced  transitions  across the minigap, and the energy relaxation of Andreev level populations. 

To this end, we couple a NS ring to a superconducting resonator, and phase bias it with a dc Aharonov Bohm flux and a small ac flux $\delta\Phi_\omega$ at the resonator's eigenfrequencies $\omega$. The linear current response $\delta I_\omega$ is characterized by the complex susceptibility $\chi (\omega)=  \delta I_\omega/\delta\Phi_\omega=\chi'(\omega) + i  \chi''(\omega) = i\omega Y(\omega)$, where $Y (\omega)$ is the NS ring's admittance. This susceptibility is extracted from the variations of the resonator's eigenmodes (frequency and quality factor). The ring's dissipationless response is deduced from the periodic flux variations of  $\chi'$, whereas the dissipation corresponds to  $\chi''$.

% can be deduced from the investigation of the current  dissipative response to a  time dependent Aharonov Bohm flux. The purpose of the work  presented here is to show how these  supercurrent fluctuations can be actually  deduced from the investigation of the frequency dependent current response of a phase biased NS ring.
%This is done by  inserting an NS ring  in a superconducting resonator submitted to a dc Aharonov Bohm  flux $\Phi_{dc}$ with  a small  ac modulation $\delta \Phi_{\omega}\cos(\omega t)$. The analysis of the perturbation  of the resonator's eigen frequencies  gives rise to the  phase dependent complex susceptibility $\chi (\omega)= i\omega Y$ where $Y (\omega)$ is the admittance of the NS ring. 

A first experiment \cite{chiodi2011}  found a large dissipative response as well as a non dissipative one that differed notably from the adiabatic susceptibility, the simple flux derivative of the ring's Josephson current (the inverse kinetic inductance). These results were partially explained by the theory of the proximity effect \cite{virtanen}.  However, the  shape of  the flux dependences  of $\chi$ did not vary with frequency (in the range explored), so that the different components of the ring's dynamical response could not be accessed. In particular, with the inelastic scattering rate  much smaller than the lowest eigen-frequency, the dissipative response associated to the relaxation of Andreev states could not be detected. 

In the present experiment, we report on a NS ring with  enhanced  temperature dependent  inelastic scattering rate $1/\tau_{in}(T)$, thanks to a thin Pd layer at the NS interface, between the normal gold mesoscopic wire and the superconducting niobium loop.
% part closing the ring.
 The higher inelastic scattering rate, combined to a  broader frequency and temperature range, lead to the identification of the two fundamental contributions to the supercurrent relaxation. At  frequencies above the inelastic scattering rate, dissipation is due to microwave-induced excitations across the minigap. In the opposite regime of lower frequency, which could not be reached previously, dissipation is due to the relaxation of Andreev level populations.  

 Here we reveal this second contribution, proportional to the sum of the squared  Andreev level currents. Accordingly we measure a response whose  period  in flux   is nearly $ \pi$ periodic. % half that of the Josephson current. 
 The extra cusps we find at odd multiples of $\pi$ reflects the closing of the minigap. This characteristic phase dependence, which is precisely  that of  the  low frequency,  thermal  supercurrent noise, is in complete  agreement with theoretical predictions formulated  long ago \cite{limpitsky,martin96,virtanen}.

\begin{figure}
\centering
\includegraphics[width=\linewidth]{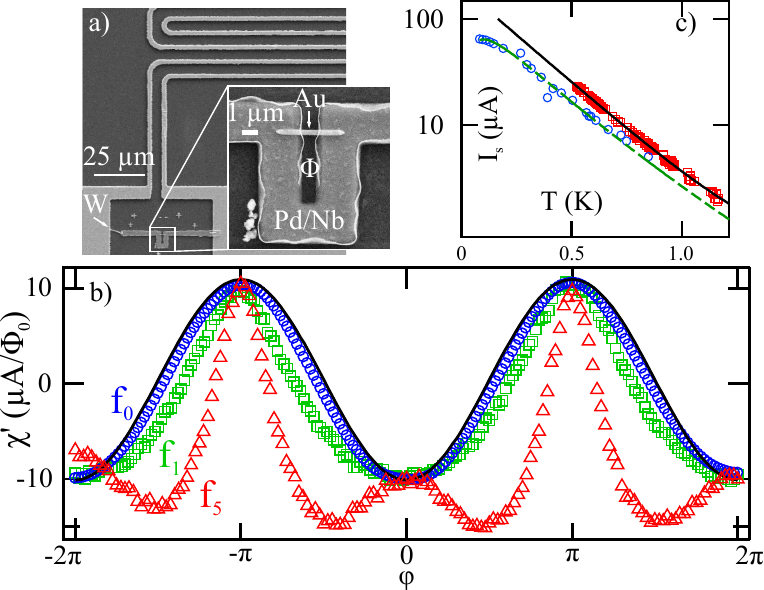}
\caption{
 \textbf{a.} Inset : Scanning  electron micrograph of the NS ring fabricated to explore the dynamics of Andreev States. \textbf{a.}NS ring is connected to a Nb multimode resonator (meander in the top) by  W wires deposited by a FIB.
\textbf{b.} Flux dependence of $ \chi ' $ at several resonator eigenfrequencies ($ f_0 = 190\,MHz,f_1=560\,MHz,f_5=2\,GHz $)  and $ T=1.2\,K $. At $ f_0$, $ \chi ' $ is barely distinguishable from a cosine (solid line).  Whereas  the amplitude $\delta\chi'_{\pi-0}= \chi '(\pi) -\chi '(0)$   is frequency independent, its temperature dependence is that of the switching current $ I_s$. Note the appearance of a local maximum around $ \varphi=0 $ at high frequency.
\textbf{c.} Temperature dependence of the  switching current of the control sample $ I_s $   (circles) and of $\frac{\Phi_{0}}{2\pi}\frac{\cal{L}}{L_C^2} \delta\chi'_{\pi-0}$  measured in the  NS ring (squares) with their  fits according to ref.\cite{dubos} (respectively dashed and solid lines). Best fit yields $ E_{Th}\equiv 71\,mK  \pm 5mK$ corresponding to $1.5\,GHz $ for the NS ring.
}
\label{fig1}
\end{figure}
The experimental set-up is shown in Fig.1a, the resonator consists in a double meander line etched out of a 1 micron thick niobium film sputtered onto a sapphire substrate. The NS ring  connects the two lines at one end of the resonator, turning it into a $\lambda/4$ line with a fundamental frequency of 190 MHz, and harmonics 380 MHz apart. A weak capacitive  coupling to the microwave generator preserves the high quality factor of the resonances, which can reach 5 $10^4$ up to 10 GHz.
The NS ring is fabricated by electron beam lithography. The Au wire  (4 micron long, 0.3 micron wide and 50 nm thick) is first deposited by e-beam deposition of high purity gold. The S part is  deposited in a second  alignment step  by sputtering of a Pd/Nb bilayer (6 nm Pd, 100 nm Nb). The resulting uncovered length of the Au wire is 1$\mu m$. The ring is connected to the Nb resonator in a subsequent step, using ion-beam assisted deposition of a tungsten wire in a focused ion beam (FIB) microscope. This process creates a good superconducting contact between the resonator and  the Pd/Nb part of the ring. The 6 nm-thick Pd buffer layer ensures a good transparency at the NS interface, as demonstrated by the amplitude of the critical current measured with dc transport measurements on  control SNS junctions fabricated simultaneously (Fig.1c). It also  enhances  the inelastic scattering rate  because of Pd's spin-wave like excitations (paramagnons) \cite{raffy,dumoulin,elke}.  Considering that the phase coherence time extracted from weak localization measurements  on a 6nm thick Pd thin film \cite{raffy}, was of the order of $0.3\pm 0.1$ ns at 1 K, which is  longer than the  estimated diffusion time $\tau_D$= $0.1ns$ through the Au wire between the S contacts, we do not expect a reduction of the critical current as confirmed by measurements  in the control samples.

The quantities we measure are  the variations with dc flux of the resonator's quality factor and eigen-frequencies  $\delta Q(\Phi)$ and $\delta f(\Phi)$.  They are simply related to the oscillating phase dependent part of the complex susceptibility, characterized by $ \chi' (\varphi)$ and $\chi'' (\varphi)$, where the superconducting  phase $\varphi$ is related  to the  flux threading the ring by  $\varphi= -2 \pi\Phi/\Phi_0$ where $\Phi_0=h/2e$ is the superconducting flux quantum. The relation reads \cite{chiodi2011}:

\begin{equation}
%\begin{array}{l}
\displaystyle\frac{\delta f_n(\Phi)}{f_n} = -\frac{1}{2}\frac{L_C^2}{\cal{L}}\chi'(\varphi),  \mbox{   }
\displaystyle\delta\frac{ 1}{Q_n }(\Phi) =\frac{L_C^2}{\cal{L}}\chi''(\varphi).
%\end{array}{}
 \label{perturb the coupling}
 \end{equation}
The coupling inductance $L_C=9\pm 2pH $ is due to the S part of the N/S ring;  $\cal{L}=$0.3 $\mu H$ is the inductance of the resonator. 
These expressions are valid at temperatures such that the  kinetic inductance of the SNS junction is larger than the ring's geometrical inductance (outside this range screening of the applied flux, both dc and ac, needs to be considered  \cite{chiodi2011}). This sets the lower limit to the temperature, so that experiments were conducted between 0.4 and 1.5 K.
The frequencies probed ranged between 190 MHz and 3 GHz.

We find spectacular variations of both the amplitude and shape of $\chi'$ and $\chi''$ as frequency and temperature are changed.
At the lowest frequencies and highest temperatures investigated (see Fig.1 and Fig.2a) the dissipationless $\chi ' (\varphi)$ is well described by a pure $2\pi$-periodic cosine, as expected for the adiabatic  susceptibility  $\chi_J= \partial I_J/\partial\Phi$ of the Josephson current which is purely sinusoidal at these moderately high temperatures, much larger than  $E_{Th}$ \cite{heikkila,strunk}. As shown in Fig.1b,1c,  The amplitude $\delta\chi'_{\pi-0}= \chi '(\pi) -\chi '(0)$  perfectly reflects the expected, roughly exponential, decay of the Josephson critical current $ I_J(T) = I_J(0)\exp(-k_BT/3.6 E_{Th})$ \cite{dubos}, that was also measured in the control wire. We find for both samples $ E_{Th}= 71 \pm 5mK$ which corresponds to $\tau_D \simeq 0.1ns$. In contrast, the dissipation, characterized by $ \chi'' (\varphi)$, is nearly $\pi$ periodic (see Fig.2b) at the largest temperatures investigated and acquires a strong $2\pi$ periodic component at lower temperature.
When increasing the frequency,  (Fig.1b and Fig.2a and 2c ) $ \chi ' (\varphi)$ contains additional harmonics, with  peaks at odd multiples of $\pi$ and  moreover a local maximum at $\varphi =0$,  mod $[2\pi]$ for the highest temperatures.  On the other hand,  at 2 GHz  (Fig2.c and 2.d ) and low temperature,   $ \chi'(\varphi)$ and $ \chi''(\varphi)$ have identical shapes, with peaks at $\pi$, mod $[2\pi]$, reflecting the underlying minigap that varies like $|cos(\varphi/2)|$.  
%Finally at the largest frequencies investigated corresponding to $hf >k_BT$ the phase dependence of $\chi'$ exhibits sharp dips around $\pi$ which are accompanied by the presence of bumps in $\delta\chi''(\phi)$ around $\phi=0$, $[2\pi]$.

\begin{figure}
\centering
\includegraphics[width=\linewidth]{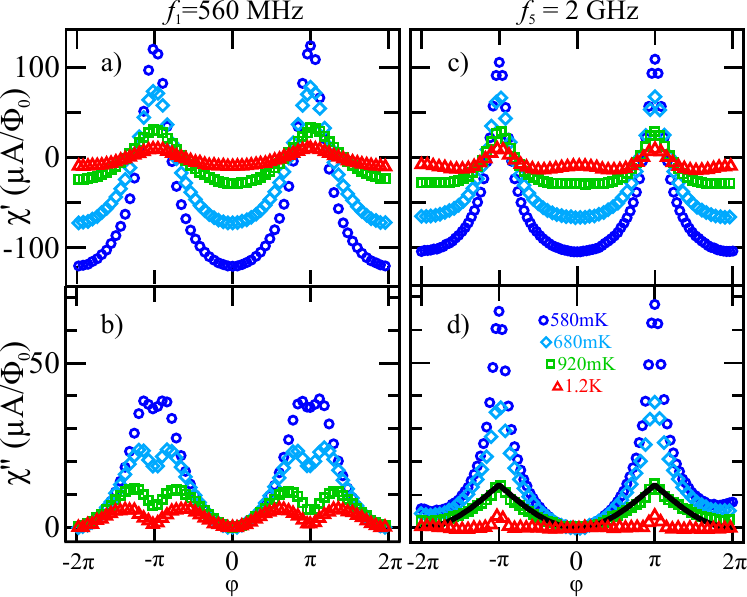}
\caption{
Evolution of $ \chi ' $ (top) and $ \chi " $ (bottom) phase dependence with temperature at $ f_1=560\,MHz<E_{Th}/h $ (left) and $ f_5=2\,GHz\gtrsim E_{Th}/h$ (right). %est-ce que ça induit en erreur de parler ici de tau_D ? ($ T=580\,mK,680\,mK,920\,mK,1.2\,K $) 
 $ \chi '(\varphi=\pi)$ and  $ \chi "(\varphi=\pi)$ increase with decreasing temperature. $ \chi '(\varphi) $ and $ \chi "(\varphi)$  strongly differ at low frequency and high temperature whereas they are similar with a shape reminiscent of the minigap at high frequency. %and low temperature. 
The solid line in d is proportional to  the minigap  dependence $|cos(\varphi/2)|$. Curves have been shifted so that $ \chi '(\varphi=\pi)-\chi '(\varphi=0)=0 $ and $ \chi "(\varphi=0)=0 $
}
\label{fig2}
\end{figure}
 In the following we exploit this complex evolution of $\chi(\varphi)$ with frequency and temperature to extract the different mechanisms  at work in the dynamics of Andreev states. To this end we make use of theoretical predictions  \cite{virtanen} based on Usadel equations, and recent numerical simulations \cite{dassonnevilletheo} inspired by the analysis of the ac response  of normal mesocopic rings \cite{buttiker,trivedi,reulet}. The response function of a NS ring has been shown to contain three contributions: $\chi =\chi_J + \chi_D +\chi_{ND}$. The adiabatic, zero frequency, Josephson contribution $\chi_J$ %-2\pi/\phi_0 
  is purely real and is the derivative of the Josephson current $\partial I_J/\partial\Phi$ . The second contribution, the diagonal susceptibility $\chi_D$, is the first non adiabatic, frequency dependent  contribution. It describes the  Debye-like relaxation  of the (phase dependent) thermal populations $f_n(\varphi)$ of the Andreev states, with a typical inelastic relaxation time $\tau_{in}$ according  to the simple model proposed  for the dynamics of persistent currents  in normal rings\cite{buttiker,trivedi,reulet}:
 \begin{equation}
 \begin{array}{l}
  \chi_D= \displaystyle \sum_n i_n \frac{\partial f_n }{\partial \Phi }\frac{i\omega}{ 1/\tau_{in} -i\omega}=  \\-\displaystyle\sum_n i_n^2 \frac{\partial f_n }{\partial \epsilon_n }\frac{i\omega}{ 1/\tau_{in} -i\omega},
  \end{array}{}
  \label{chiD}
 \end{equation}
 where the square of $i_n = -\partial\epsilon_n/\partial\Phi$,  the current carried by the n-th Andreev level of energy $\epsilon_n(\Phi)$, appears. Finally, the non-diagonal contribution $\chi_{ND}$ describes quasi-resonant  microwave-induced transitions between two Andreev levels, involving  (in contrast with $\chi_D$) non diagonal matrix elements of the current  operator \cite{dassonnevilletheo}. 
The  contribution  $\chi''_{ND}$ to the phase dependent susceptibility dominates  when $\omega \geq 1/\tau_D \gg 1/\tau_{in}$, as in Fig.2d. $\chi'$ and $\chi''$   then  have similar  shapes which follow approximately the minigap  with peaks at $\pi$ and a  $|cos(\varphi/2)|$ dependence\cite{dassonnevilletheo}. This high frequency regime   was the only one accessed in the previous experiments on Au wires directly connected to W superconducting wires.  In  those experiments the energy relaxation time, limited by electron electron interactions,  of the order of $0.1\mu s$,  was very long due to the superconducing contacts\cite{chiodi2011,blanter96}. Therefore those measurements were always in the regime $\omega \tau_{in} \gg 1$ where $\chi''_D$ is negligible, (Eq.\ref{chiD}). In contrast, the Pd layer beneath the Nb contacts in the present samples considerably reduces the inelastic scattering time, leading to a substantial contribution of $\chi''_D$ for the  resonator's first five eigenfrequencies. We now  focus on this contribution analyzed in Fig.3 and Fig.4.
%, and to leave a more detailed analysis of $\chi_{ND}$ and its comparison with the minigap for future studies.

 %In order to compare our experiments with theory we consider the quantity describing the predicted dependence of $\chi_D$: 
 We first present the predicted flux dependence of $\chi_D$, given by the function F
 \begin{equation}
  F(\Phi,T)=\sum_n i_n \frac{\partial f_n }{\partial \Phi }= -\sum_n i_n^2 \frac{\partial f_n }{\partial \epsilon_n },
  \label{F}
 \end{equation}
 which reads in the continuous spectrum limit $F(\Phi,T)= \int d\epsilon J_S^2(\Phi,\epsilon)/\left[4k_BT\rho(\epsilon) \cosh^2(\epsilon/2k_BT)\right]$. Here $J_S$  and $\rho$ are respectively the spectral current and the density of states of the SNS junction. This function was  introduced by Lempitsky \cite{limpitsky} to describe the I(V) characteristics of  SNS junctions, and was calculated numerically using Usadel equations by Virtanen et al.\cite{virtanen}. At large temperature compared to $E_{Th}$, $F(\Phi,T)$ can be approximated by the following analytical form: $F_U(\varphi,T)\propto \left[(-\pi + (\pi + \varphi)[2 \pi])\right]\sin (\varphi) - |\sin (\varphi)|\sin ^2(\varphi/2)/\pi  $. It is dominated by its second harmonics  with in addition a sharp linear singularity at odd multiples of $\pi$ (see Fig.4). This is due to the dominant contribution  of Andreev levels close to the minigap whose flux dependence  is singular like in a highly transmitting superconducting  single channel point contact \cite{martin96}.  %As also noted in this context of point contacts, this  singularity is enhanced  by the derivative of the Fermi function leading to  a sharp  dip at low temperature. 
  \begin{figure}
  \centering
  \includegraphics[width=1\linewidth]{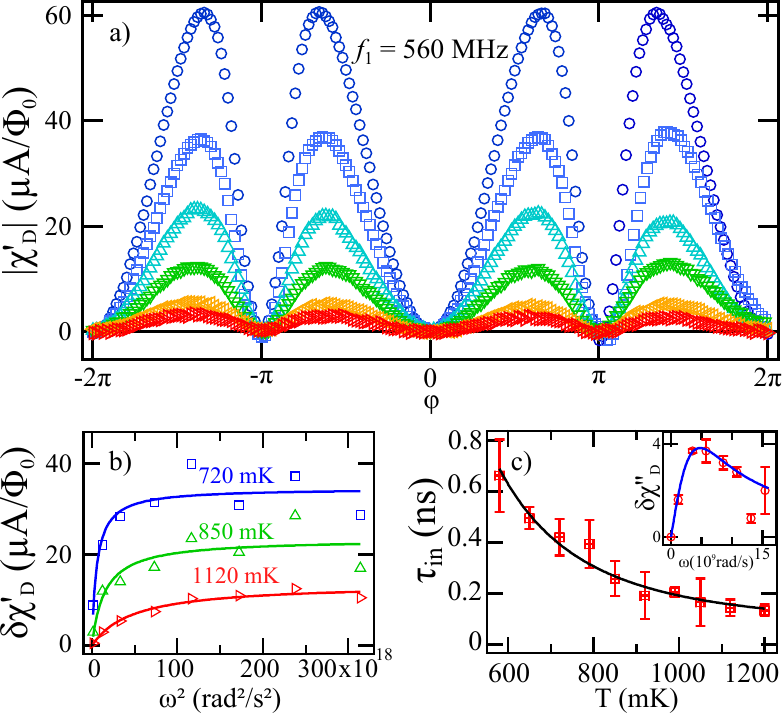}
  \caption{
  {\bf a.} Experimental  phase dependence of $ \chi'_D $ at $ f_1=560\,MHz $ and (from top to bottom) $ T=580\,mK,$ $ 650\,mK,$ $720\,mK,$ $820\,mK, 1000\,mK,$ $1200\,mK$. In  agreement with theory \cite{virtanen}, the position of the maximum slightly shifts to lower values with decreasing temperature.
 {\bf b.} Frequency dependence of  $ \delta \chi'_D$ :  the maximum  of $-\chi'_D(\varphi)$, at different temperatures (triangles) compared to the theoretical prediction from Eq.3. {\bf c.} Temperature dependence of extracted $ \tau_{in} $ (see text) . Solid line is a $ T^{-3} $ law. Inset: frequency dependence of $ \delta \chi''_D$ at 1.15K (circles) compared to the theoretical prediction from Eq.3.
 %At a given frequency $ \chi'_D $ amplitude,$\delta \chi'_D $, is decreasing with temperature and the position of the maximum slightly changes from $  0.36\,\phi_0$ at $ T=580\,mK $ to a saturating  $  0.3\,\phi_0$  for $ T>800\,mK $. Since  $ \chi'_D $ frequency dependence is only due to $(\omega\tau_{in})^2/(1+(\omega\tau_{in})^2)$  , fitting $\delta \chi'_D (\omega^2)$ at different temperature yields $ \tau_{in} $ temperature dependence. 
  }
  \label{fig3}
  \end{figure}
 We now show that the particular flux dependence of the function $F_U$ can explain the experimental data of  Fig.1 and 2.  
We first follow the frequency dependence of the   amplitude  of  $ \chi'_D(\varphi)=\chi'(\varphi) - \chi_J(\varphi)$  at fixed temperature,  and check that the shape of % the phase dependence of
 $ \chi'_D(\varphi)$ does not change with frequency and is the same as that of $ \chi''_D$, as predicted for the temperature  and  frequency regime  where the contribution of $ \chi''_{ND}$ can be neglected.   As shown on Fig.3b it is then possible to  fit the frequency dependence of the amplitude of $\chi'_D(\varphi)$ by the expected $(\omega\tau_{in})^2/(1+(\omega\tau_{in})^2)$  and determine the characteristic time $\tau_{in}$ for several temperatures  according to Eq.\ref{chiD} . We find values of $\tau_{in}$  varying between $0.2$ and $1$ ns, quite similar  to what was deduced from weak localisation measurements  in Pd thin films \cite{raffy}. Moreover  the power law decrease of $\tau_{in}(T)$ in $T^{-3}$ , is in reasonable agreement with  what is expected for paramagnons which constitute  the dominant inelastic scattering at low temperature in    Pd which is close to a ferromagnetic transition. It is also  interesting to note that our results can be described by a single inelastic time,  independent of $\varphi$, whereas a phase dependent $\tau_{in}$ is expected for electron phonon collisions in SNS junctions \cite{giazotto}. This is probably due to the fact that temperature is larger than $E_g(0)$ in our case. 

 A similar analysis can be done  on $\chi''$ ,  the quality of the calibration is however not as good as on $\chi'$. Moreover we still  lack a good analytical prediction for $\chi_{ND}(\varphi)$,  which gives a large contribution to $\chi''(\varphi)$ at low temperature and high frequency.  We have overcome this difficulty by subtracting for frequencies larger than 1.7~GHz the  flux dependence of $\chi''_{ND}$ estimated from the high frequency data (2.8~GHz).   The resulting amplitude $\delta\chi''_D (\omega)$   agrees with the expected frequency dependence in $\omega\tau_{in}/(1+(\omega\tau_{in})^2)$ as shown in the inset of  Fig.3c. One can also compare the independently  measured  flux dependences of  $\chi'-\chi_J$ and $\chi''$    with theoretical predictions from the Usadel equations,  $F_U(\Phi,T)$. This is done  in Fig.4 for several frequencies and a good agreement is found.
 \begin{figure}
     \centering
    \includegraphics[width=\linewidth]{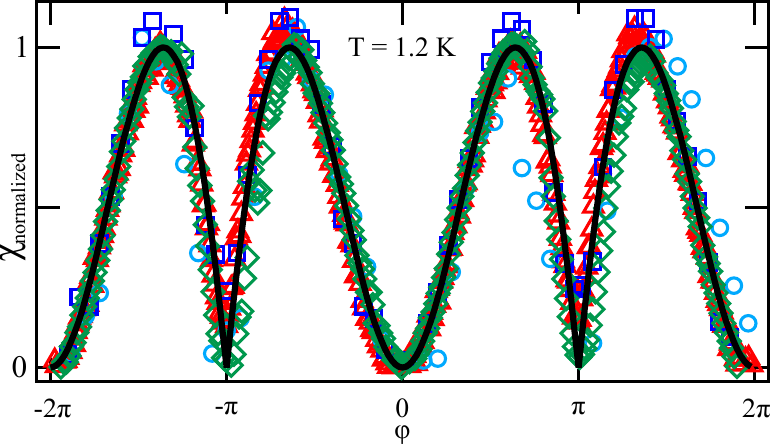}
     \caption{ %mettre ça à 2 fréquences différentes
 Normalized experimental  phase dependences at $ T=1.2\,K $ (open symbols) compared to $ F_{U} $ (solid).
 Data are $\chi" $ (squares),  $|\chi'_D|=\chi_J -\chi'$ (circles) at $ f_1=560\,MHz $ and $\chi" $ (triangles), $|\chi'_D|=\chi_J -\chi'$ (diamonds) at  $ f_3=1.35\,GHz $. At high temperature and for  $ f \lesssim E_{Th}/h $, experimental $\chi_J -\chi'$ and $ \chi" $ are found to be in very good agreement with theoretical prediction of Usadel equations $ F_{U} $  . 
      }
      \label{fig4}
     \end{figure}
		
 With this set of experiments, we have thus shown that the frequency and temperature dependences of the response function of NS rings in a time dependent flux are consistent with a simple  Debye  relaxation model of the population of the Andreev levels. Using fluctuation dissipation theorem one can estimate the related thermodynamic current noise as:
\begin{equation}
\displaystyle   S_I(\omega) =\frac{2}{\pi} \frac{k_B T\chi''_D(\omega)}{\omega}=\frac{2}{\pi} k_B T \sum _n  i_n^2(\varphi)\frac{\partial f_n}{\partial\epsilon_n }\left[\frac{\tau_{in}}{1+(\omega\tau_{in})^2}\right].
\end{equation}

 The measurement of the ac current linear response of an NS  ring to an ac flux thus reveals two fundamental mechanisms contributing to dissipation at finite frequency.  One of them, predominant at high frequency and low temperature,  describes the physics of microwave induced transitions  above the minigap.  We have clearly identified and characterized the second cause of dissipation, the  thermal relaxation of the populations of the Andreev states. It is  described by  an inelastic rate which is extremely sensitive to the nature of the NS interface. This dissipative response  is directly related to the low frequency thermal noise of the Josephson current, with a flux dependence proportional to the average square of the spectral (or single level) current, and can be precisely described by theoretical predictions. These results  show that linear ac measurements in a wide range of frequency, close to equilibrium, reveal physical properties of SNS junctions that are not accessible by standard transport measurements dominated by non linear effects. The type of  experiments presented here is uniquely suited to investigate more exotic systems, for instance with the normal diffusive wire replaced by a  ballistic  wire,  leading to  a discrete Andreev spectrum known to be extremely sensitive to spin orbit interactions.
 
 We acknowledge A. Kasumov and F. Fortuna for help with the FIB and  M. Aprili, F. Chiodi, R. Deblock, M. Feigelman, T.T. Heikkil$ \ddot{\rm a} $, K. Tikhonov  and  P. Virtanen for fruitful discussions.

\end{document}